\documentclass{aa}
\usepackage{graphicx}
\usepackage{txfonts}
\begin{document}
   \title{On  the  new   short  orbital  period  cataclysmic  variable
   H$\alpha$0242-2802\thanks{Based on observations obtained at the ESO
   Paranal Observatory (ESO proposal 269.C-5056(A)).}}

   \subtitle{   }

   \author{E. Mason
          \inst{1}
          \and
          S. B. Howell\inst{2}
          }

   \offprints{E. Mason}

   \institute{ESO, Alonso de Cordova 3107, Casillas 19001, Vitacura
              Santiago, Chile\\
              \email{emason@eso.org}
         \and
             WIYN/NOAO, P.O. Box 26732, 950 Cherry Ave., Tucson, AZ 85719\\
             \email{howell@noao.edu}
             }

   \date{Received September 15, 1996; accepted March 16, 1997}

   \abstract{ We present  results from our phase-resolved spectroscopy
of  the  newly  identified  cataclysmic  variable  H$\alpha$0242-2802.
H$\alpha$0242-2802 was  identified from a deep  UK Schmidt H$\alpha$-R
band survey as a candidate cataclysmic variable.  Although initial
spectroscopy revealed an optical spectrum  very similar to that of the
famous short  orbital period system  WZ~Sge, the analysis of  our data
shows that this is not  the case. H$\alpha$0242-28 differs from WZ~Sge
in  both the  accretion  disk  structure and  the  orbital period.  In
particular,  H$\alpha$0242-28 appears  to be  a system  which  has not
reached the orbital period minimum, yet. 
 
   }

   \maketitle
%

\section{Introduction}

  Cataclysmic variables  (CVs) are binary systems where  a white dwarf
  (WD) accretes  material from a  secondary star which is  filling its
  Roche lobe (see Warner 1995 for a review).

Recent theoretical  works (e.g.  Howell et al.   2001, and  Kolb 2001)
predict that  the majority of the  CVs should be very  old and evolved
systems. These systems are  characterized by short orbital periods
(i.e. $P<$2 hr), low  intrinsic luminosity ($M_V<11$ mag), and evolved low
mass secondary  stars (see also Howell  et al. 1995).  However, the
number of  currently known/observed short orbital  period systems does
not match  the expectation. Thus,  in recent years, many  surveys have
taken place to fill  such a gap (e.g. SDSS by Szkody  et al. 2002, HQS
by Gansicke et  al. 2002, FSVS by  Groot et al. 2003, etc).  At a
same time, new ideas (e.g., the need to go even fainter) and different
accretion scenarios  (e.g. Spruit and  Taam 2001, Dubus et  al. 2002),
have been presented and analyzed.

H$\alpha$0242-2802 (hereafter H$\alpha$0242) was a CV candidate within
the UK-Schmidt survey  (Davenhall et al. 2001). It was confirmed
as CV by Howell et  al.  (2002), who observed it spectroscopically and
reported  a spectrum  very  similar  to that  of  WZ~Sge. Indeed,  the
authors also advanced the hypothesis that H$\alpha$0242 belongs to the
same dwarf nova subclass  of Tremendous Outburst Amplitude Dwarf novae
(TOADs, see Howell et al. 1995 for a review on TOADs).  An interesting
implication of  this would be  that H$\alpha$0242 is quite  high above
the galactic plane  ($z=250$ pc), and resides in the  hold disk - halo
population. 

Here, we present time-series spectroscopy obtained with VLT+FORS2 with
the aim of determining the  orbital period, the system parameters, the
emission lines  characteristics, and comparing such  values with those
of WZ~Sge, the prototype object  of the very old evolved CV population
(see Howell et al.,  2004).  We present our spectroscopic observations
in Sec.~2, the data analysis in Sec.~3 and our conclusions in Sec.~4

\section{Observations and data reduction}

H$\alpha$0242  was observed  at  VLT+FORS2  using a  series  of 5  min
exposures  over  4  consecutive  hours.  An  8-m  class  telescope  is
mandatory  in order  to perform  time resolved  spectroscopy  of faint
targets (H$\alpha0242$ B mag is $\sim$19, see Howell et al. 2002), and
determine the  radial velocity curve of systems  having orbital period
around 2  hr.  We used  FORS2 in Multi  Object Spectroscopy (MOS)
mode with  the holographic grism 1400V.  We preferred the  MOS mode to
the long slit spectroscopy (LSS), in order to cover a bluer wavelength
range\footnote{Namely  $\lambda\lambda$  4270-5550,  rather  than  the
standard  $\lambda\lambda$4560-5860.}.  Indeed,  the  bluer  wavelength
coverage  allowed us  to observe  the Doppler  broadened  Balmer lines
H$\beta$ and H$\gamma$  (see Fig.~1).  The slit width  was set to 1''
and  together with  the  grism  1400V provided  a  dispersion of  0.63
\AA/pix.

   \begin{table}
         \label{log}
 \begin{center}
\scriptsize
\caption{The log. of observation.}
\begin{tabular}{cc}
date of obs. & 2002/09/10 \\
UT start &  05:41\\
UT end & 09:52 \\
average seeing & 0.7  \\
average transparency  & CLR-THN\\
number of spectra & 40 \\ 
exptime per spectrum &  300 sec \\
instrument & FORS2\\
instr. set up & MOS/SR/no filter/GRIS 1400V\\
read out mode & 200kHz, 2x2, 1.25 (speed, binning, gain)\\
\end{tabular}
\end{center}
   \end{table}

   \begin{figure*}
   \centering
   \rotatebox{-90}{\includegraphics[width=14cm,]{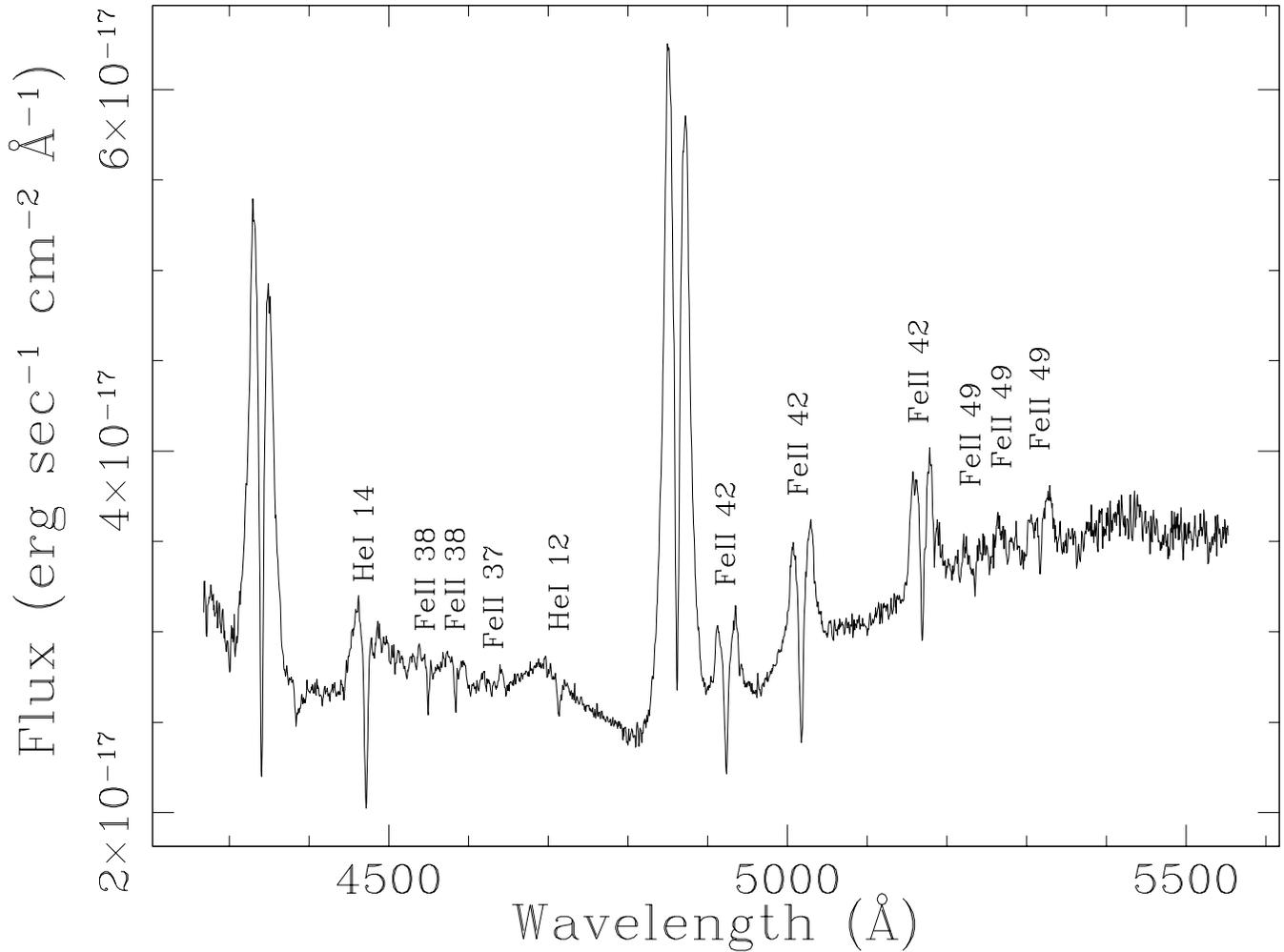}}
      \caption{The  average  spectrum of  H$\alpha$0242.  The flux  is
given in erg sec${-1}$ cm$^{-2}$  \AA$^{-1}$, but has not been slit or
sky corrected thus does not correspond to an absolute calibration.}
         \label{f1}
   \end{figure*}

The log.  of observation  is presented in  Table~1 .  A  standard star
with the same instrument set  up was observed one month later allowing
us  to place  the  spectra  on a  relative  flux scale.   Fortunately,
absolute flux  measurements are not needed for  the analysis presented
in this paper.

The data reduction was performed through standard IRAF routines within
the packages {\it ccdproc}, {\it apex}, and {\it onedspec}.
 
\section{Data analysis}

\subsection{Overview/General properties}
We plot  in Fig.1  the average  spectrum of  H$\alpha$0242.  The
continuum  shape is  not as  blue as  in the  previous  observation of
Howell  et  al.   (2002) and  this  is  probably  an artifact  of  the
calibration.  The emission lines which are visible in the spectrum are
the two  Balmer lines H$\beta$ and  H$\gamma$ and the  FeII lines from
multiplets 37,  38, 42 (the strongest),  and 49. Visible  are also the
HeI lines $\lambda$4471 and $\lambda$4713.  The observation of optical
FeII emission  has already been  reported in the literature  but never
with sufficient  attention.  We will develop a  brief discussion about
FeII emission and their formation mechanism in CVs in Sec.~3.6.  Here,
we note  that all the emission  lines have a similar  shape.  They are
double peaked with a deep  central absorption core which extends below
the continuum.   This is signature  of high orbital  inclination.  The
B/R ratio of the {\it  average spectrum} (Fig.~1) appears greater than
1 for the Balmer lines and $<$1  for the FeII. This is, at least in
part,  an artifact  of the  flux  calibration which  introduces a  red
continuum.  Indeed, on one hand, the {\it normalized average spectrum}
shows that B/R$>$1 in the Balmer lines and B/R$\sim$1 in the FeII (42)
emission lines.   On the other hand,  the analysis of  the {\it single
spectra}  shows that  the  Balmer, HeI,  and  the FeII  lines follow  a
similar modulation  throughout the orbit.   This is well shown  by the
trailed  spectrograms that we  present further  below (see  Fig.~5). A
possible explanation  for the different  B/R ratio between  the Balmer
and the  FeII lines in  the {\it average  spectrum}, may be  a larger
fractional contribution of  the hot spot in the  Balmer lines, when it
is blue shifted.

We also observed a weak emission from the HeII $\lambda$4684. The HeII
line is not  readily  visible in  the  single spectra,  nor in  the
average one.  It is visible  only in the trailed  spectrogram where
it produces a pure S-wave (see Fig.~5 and Sec.~3.4).

In the  following sections we  will analyze the spectra  searching for
the  orbital period  of  the  binary system,  and  applying the  usual
analysis  for  the accretion  disk  emission  lines (RV  measurements,
trailed  spectra  etc).   We  will  pay particular  attention  to  the
comparison between WZ~Sge and H$\alpha$0242. 

\subsection{Period search}

   \begin{figure*}
   \centering
   \rotatebox{-90}{\includegraphics[width=14cm,]{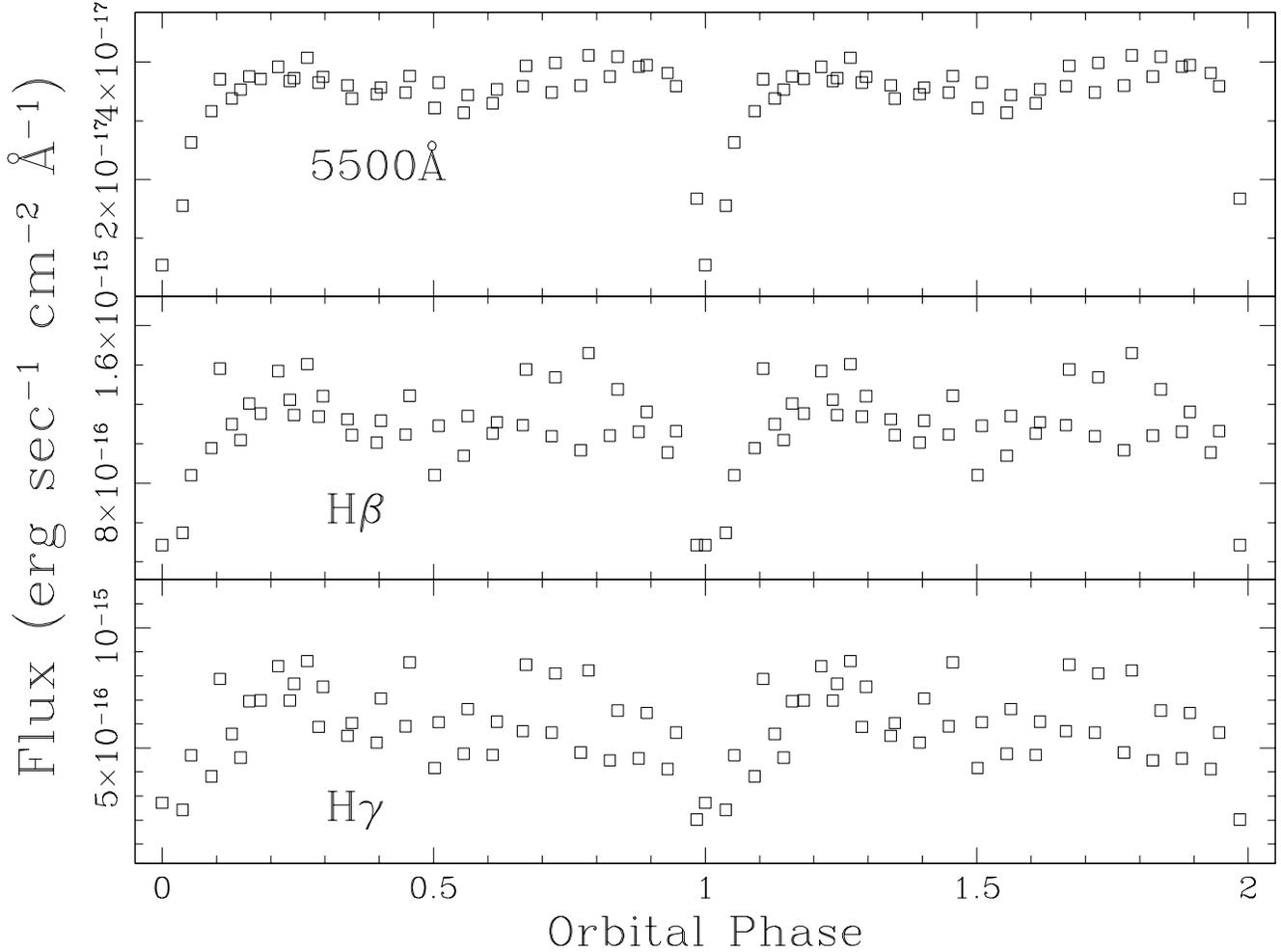}}
      \caption{The   continuum   and   Balmer  emission   line   light
      curves.  From top  to  bottom: the  continuum  flux measured  at
      5500\AA,  H$\beta$ emission lines  flux, and  H$\gamma$ emission
      line flux. }
         \label{f2}
   \end{figure*}

We  searched for  the binary system  orbital period  applying the
Phase Dispersion  Minimization method (PDM  method, Stellingwerf 1987)
to   different   emission   line  features\footnote{We   measured   in
particular: the position of the red  and the blue peak in the emission
line,  the position  of  the  central absorption,  and  the line  flux
barycenter.}.   We  did  not  discover any  statistically  significant
period. However,  we found  clear evidence for  the orbital  period by
plotting the light  curve of the continuum flux  measured at 5500\AA \
(see  Fig.~2). Indeed,  the  light  curve is  characterized  by a  deep
eclipse of almost 2 magnitudes with periodicity of 107 min.  Woudt et
al. (2004),  determined the same orbital period  through time resolved
photometry.

We also  found a periodicity of 106 min  from the radial velocity
measurements of the Balmer  emission lines through the double Gaussian
fit  method  which  was  developed  by Shafter  (1983,  and  reference
therein). 

Within this paper, we will adopt  the period of 107 min and will phase
our spectra  assuming as  $T_0$ the time  of the  observed mid-eclipse
minimum\footnote{Due  to  the  time  resolution of  our  spectra,  the
``true'' minimum could be deeper.}. Our approximate ephemeris is:

$HJD=2452527.89275(\pm 0.00395)+0.0743055(\pm 0.0017361)E  $

where  the uncertainty  on  the  time $T_0$  corresponds  to the  time
resolution of  our data points which is  341 sec, i.e. the  sum of the
exposure  time and  the  readout time  of  the mosaic  CCD. Thus,  the
uncertainty on the  period (150 sec), was derived  both from measuring
the width  of the period peak  in the standard PDM  ($\theta$, P) plot
and an estimation from the eclipse light curves in Fig. 2.

   \begin{figure}[h]
   \centering
   \rotatebox{-90}{\includegraphics[width=6.8cm]{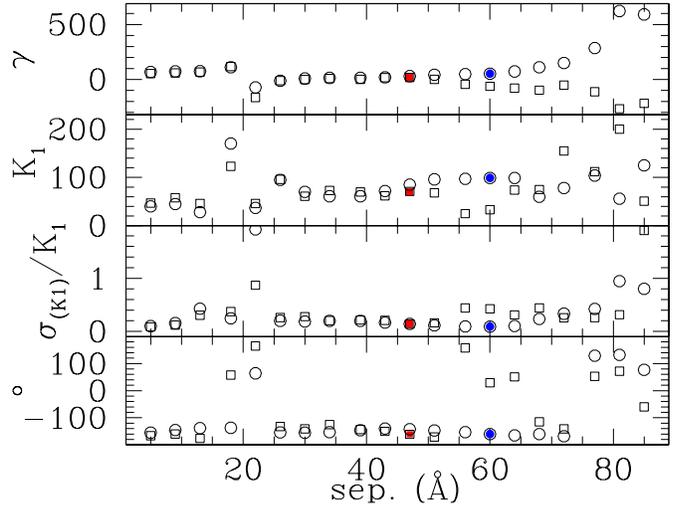}}
      \caption{The  diagnostic diagram.  Squared symbols  are  for the
      H$\gamma$  data points;  circles are  for H$\beta$  data points.
      Filled symbols mark the best fit data points.  }
         \label{f1}
   \end{figure}

\subsection{Radial velocity curves}

Knowing the  orbital period  we can  phase the  spectra  and the
radial velocity measurements to derive a few parameters characterizing
the  binary  system. We  can  then  produce  trailed spectrograms  and
Doppler  maps  to qualitatively  analyze  the  evolution  of the  line
profile and the emission components.

In  principle,  radial velocity  measurements  of  the accretion  disk
emission lines  provide valid information  on the white  dwarf orbital
motion. However, this does not appear to be the case for many CVs and,
in particular,  the short orbital period systems  (e.g.  WZ~Sge, Mason
et al.  2000 and reference therein,  and V893 Sco, Mason et al. 2001).
The exact  causes of such behavior is  not well known, nor  is there a
model capable to explain why the radial velocity curves from different
emission lines  yield discordant system parameters.  Since  we are not
yet  able  to  quantify  the  discrepancy  between  the  derived
quantities and  the corresponding real values, it  is still worthwhile
measuring and fitting the radial velocity curves of the emission lines
and  determining  coarse values  which  reasonably  constraint the  WD
Keplerian velocity  and the systemic velocity.  Of  course, within our
uncertainties,   the   readers   should   be   cautionary   in   their
interpretation. 

We measured the  radial velocity of the Balmer  lines using the double
Gaussian  fit  method as  defined  by  Shafter  (1983, and  references
therein).   We  used  narrow  Gaussian  FWHM  (4-2\AA),  and  Gaussian
separation steps of equal size. The diagnostic diagram in Fig.~3 shows
the fitting parameters  corresponding to steps of 4  \AA. The best fit
for  the  H$\beta$  radial  velocity  curve is  found  at  a  Gaussian
separation  of 60 \AA,  while the  best fit  for the  H$\gamma$ radial
velocity curve  is at a separation of  47 \AA. We list  in Table~2 the
best fit  parameters and  plot in Fig.~4  (two top panels)  the radial
velocity curves.  As expected and already observed in other short
orbital period systems, the two Balmer lines do not produce consistent
values for the WD Keplerian  velocity and the systemic velocity, which
indeed differs by  more than $3\sigma$.  In addition  we may note that
the two  radial velocity curves  consist of largely  scattered points.
We believe that one possible explanation can be found in the fact that
the line  profile is largely variable across  different orbital cycles
(see Fig.~5 and  Sec.~3.4).  Also, the larger scatter  of the velocity
measurements in  the H$\gamma$ lines  are probably due to  the smaller
Gaussian separation,  which, in  principle, can be  biased by  the hot
spot S-wave.  However,  a larger separation does not  produce a better
result due to the noisy wings of the H$\gamma$ emission line.  We will
thus adopt  a white dwarf Keplerian  velocity of 99  km/sec (from just
the H$\beta$  line) throughout the  present work.  Since  the systemic
velocity  $\gamma$ resulting by  the application  of the  Gaussian fit
method  is  the  most  un-reliable  parameter (see  Tappert  1999  for
details),  we  will adopt  the  value  of  $\gamma=22$ km/sec  derived
below. 

The  HeII  emission  is visible  only  as  an  S-wave in  the  trailed
spectrogram  (see Fig.5), thus,  the only  way we  had to  measure its
radial velocities was by cursor position and visual inspection.

   \begin{figure*}
   \centering
   \rotatebox{-90}{\includegraphics[width=14cm,]{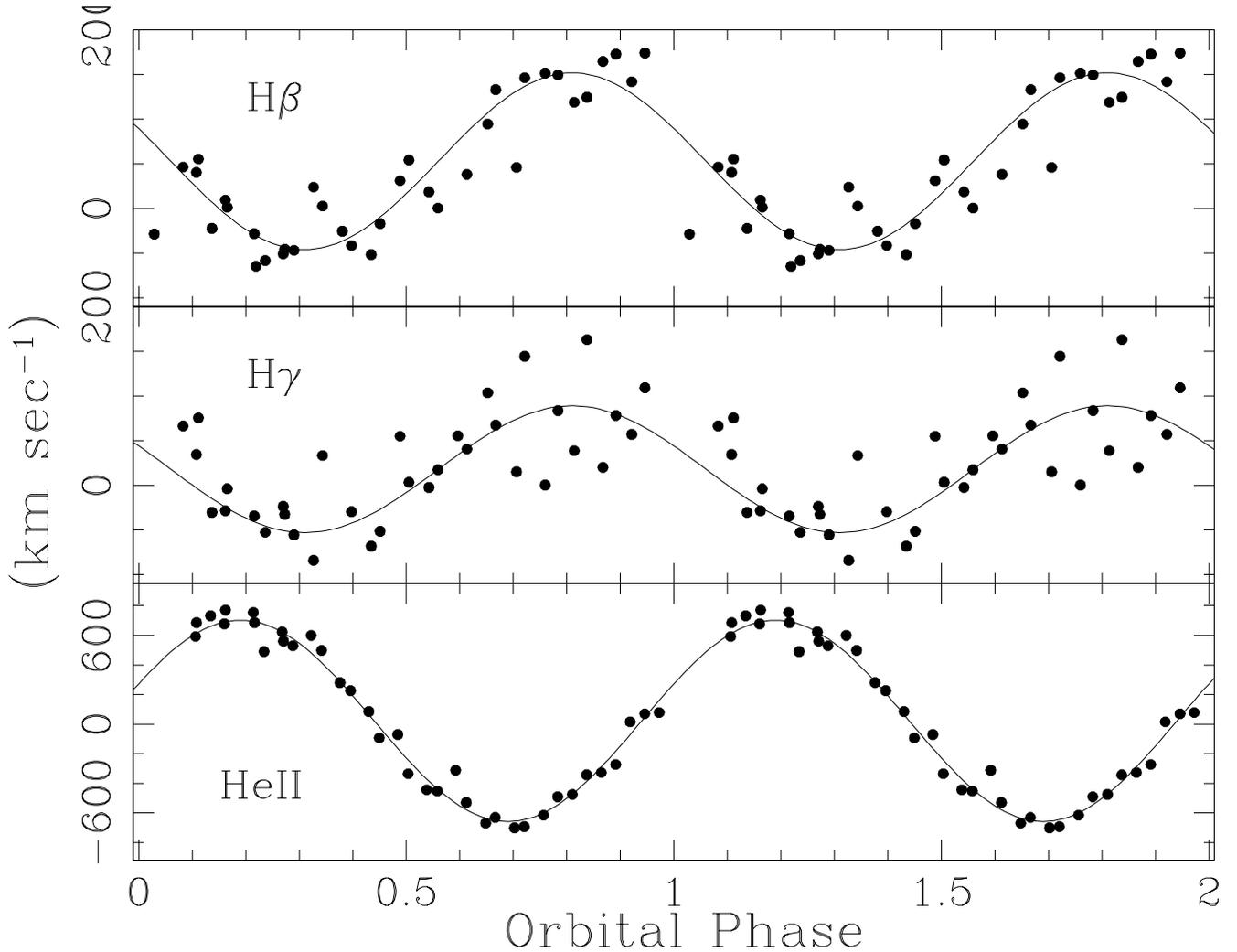}}
      \caption{The  radial velocity  curves.  From top  to bottom:  1)
      H$\beta$  accretion disk emission,  2) H$\gamma$  accretion disk
      emission, 3) HeII hot spot emission. }
         \label{f4}
   \end{figure*}

\begin{table}[h]
  \label{log}
\begin{center}
\scriptsize
\caption{Best fit  radial velocity curve  parameters for the  H Balmer
lines and HeII 4686.}
\begin{tabular}{cccc}
em. line & $\gamma$ (km sec$^{-1}$) & K$_1$ (km sec$^{-1}$) & $\phi_{R/B}$ \\
 & & & \\
H$\beta$ & 53$\pm$6 & 99$\pm$9 & 0.056$\pm$0.014 \\
H$\gamma$ & 18$\pm$7 & 71$\pm$10 & 0.053$\pm$0.025 \\
HeII & 22$\pm$3 & 680$\pm$4 & 0.44$\pm$1E-3 \\
\end{tabular}
\end{center}
\end{table}

The radial  velocity curve  is plotted in  Fig.~4, while its  best fit
parameters  are   reported  in   Table~2.  The  derived   $K_1$  value
corresponds  to  the Keplerian  velocity  of  the  outer edge  of  the
accretion  disk, to  a  first approximation.  Indeed,  the real  value
should be  larger as the WD (instantaneous)  radial velocity subtracts
off  from  each   spectrum/measurement.  However,  the  derived  value
$K_1=680$ km/sec  matches fairly well  the half peak  separation (HPS)
measured  on the  Balmer lines:  $\langle HPS  \rangle =  650  \pm 46$
km/sec. The HPS  is also a good estimate of  the Keplerian velocity at
the outer  accretion disk edge. We  used the value of  K=680 km/sec to
determine  the  accretion  disk  size.  Assuming a  WD  mass  of  0.64
$M_\odot$   (see  Sec.~3.5)  we   find  $r_d\sim   1.8\times  10^{10}$
km/sec\footnote{The  slightly smaller  velocity derived  from  the HPS
produces  a  larger  accretion  disk  radius  of  $r_d\sim  1.98\times
10^{10}$  km/sec. However,  we believe  that the  average  spectrum is
affected by  both the orbital motion  and the hot  spot emission which
reduce the peak separation.}.

The derived value of $\phi_{R/B}=0.44$ is consistent with the standard
hot  spot  position at  an  angle of  $\sim$50-60  deg  from the  line
connecting the  two star  centers of mass.   While, the  HeII $\gamma$
value  corresponds   to  just   the  systemic  velocity,   within  our
assumption. This is the value adopted in our next analysis.

\subsection{Trailed spectra and Doppler maps}

   \begin{figure*}
   \centering
   \rotatebox{-90}{\includegraphics[width=13cm,]{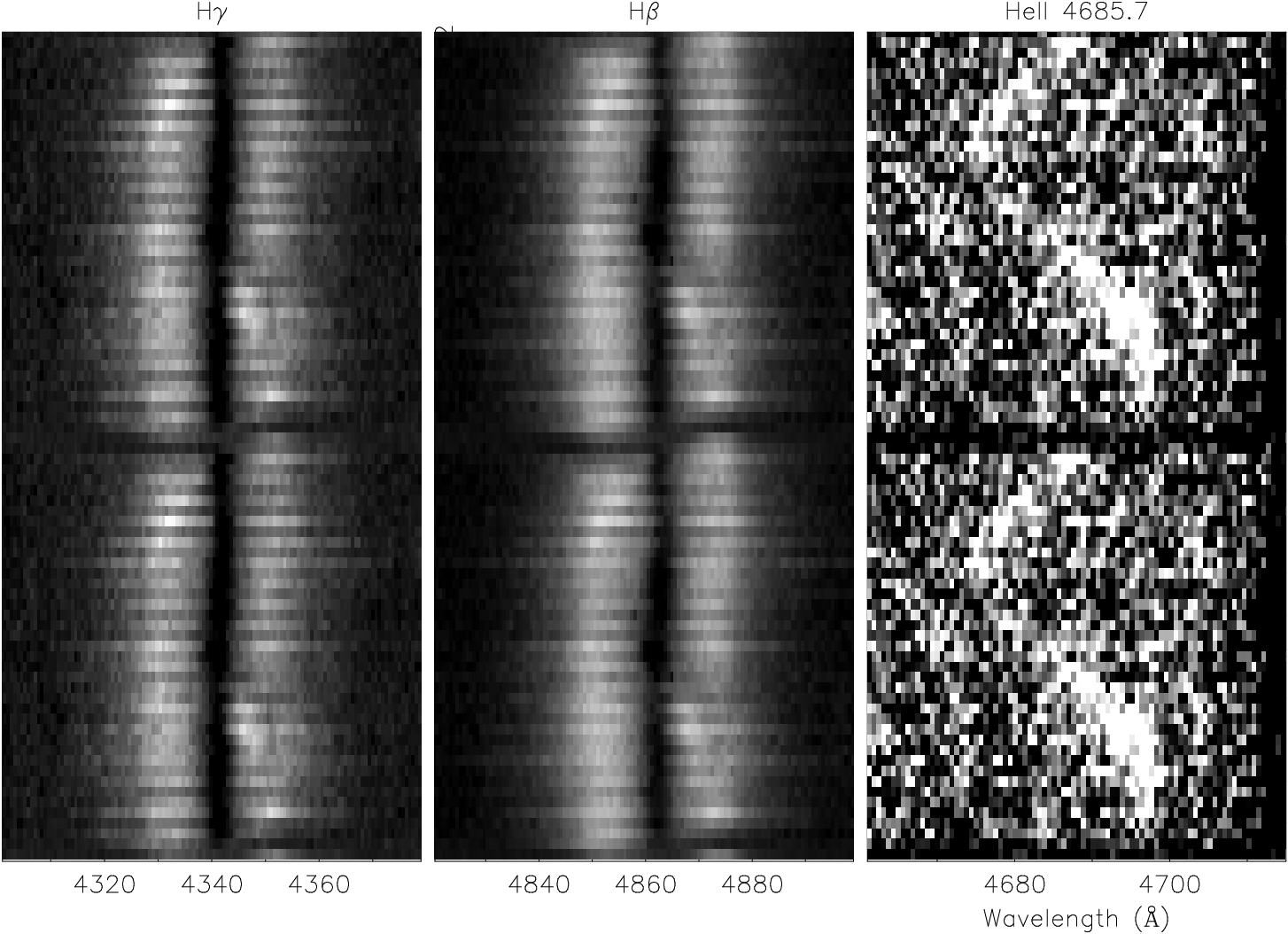}}
   \rotatebox{-90}{\includegraphics[width=13cm,]{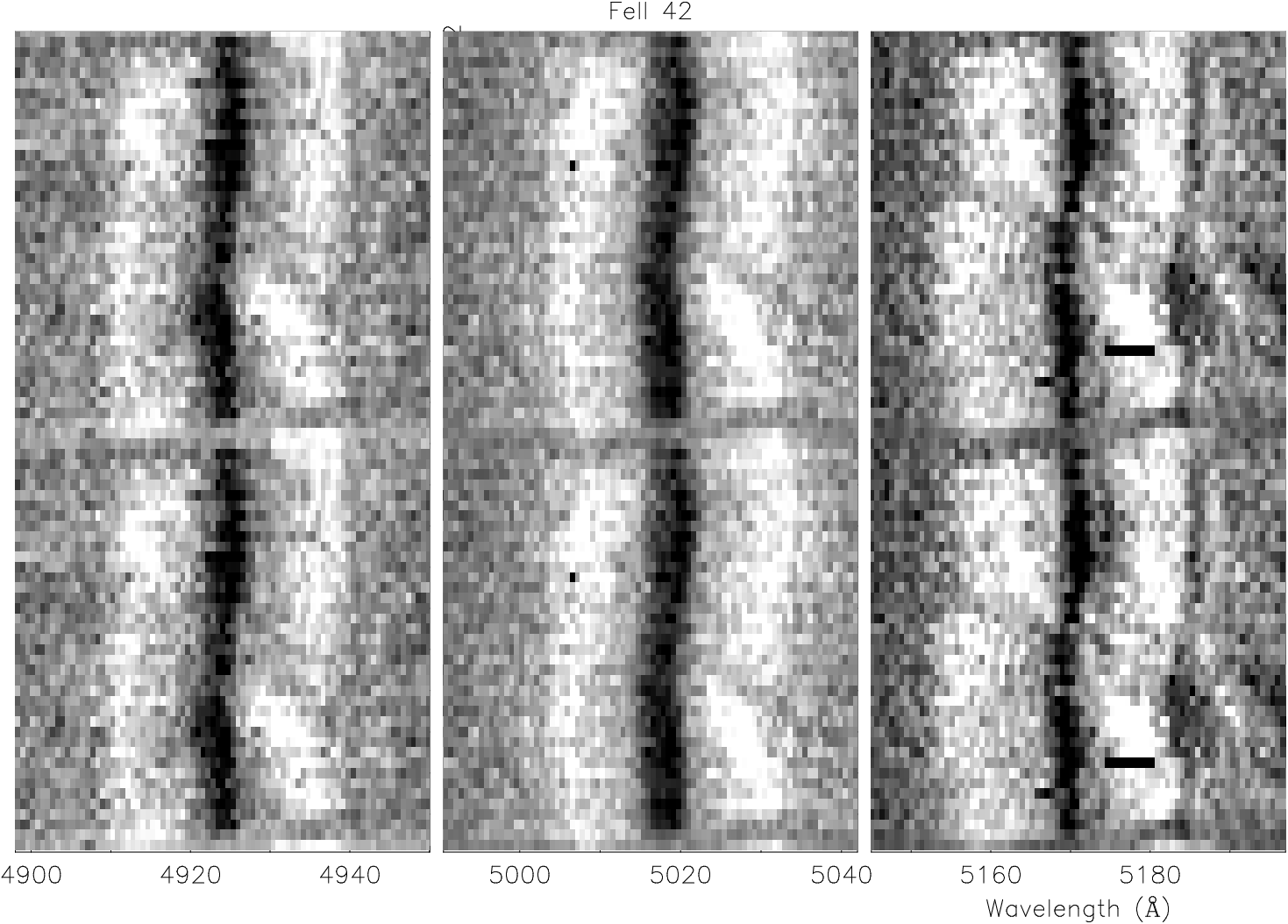}}
      \caption{Trailed  spectrograms: top  panel from  left  to right:
      H$\gamma$,  H$\beta$,  and  HeII $\lambda$4686  emission  lines;
      bottom panel: FeII 42 emission lines. }
         \label{f5}
   \end{figure*} 

Trailed  spectrograms and Doppler  maps of the  observed emission
lines  are a  common tool  to qualitatively  analyze the  line forming
region.  We thus plot in Fig.~5 the trailed spectrograms of the Balmer
lines and the HeII $\lambda$4686 (top panels of Fig.~5), as well as of
the FeII (42) lines (bottom panels of Fig.~5).  All the emission lines
clearly show  evidence for  an eclipse, which  implies a  high orbital
inclination of the  binary system (with possibly total  eclipse of the
white dwarf itself). The Balmer  and the FeII lines are double peaked,
i.e. form in the upper  atmosphere/corona of the accretion disk. They
also show only weak evidence of  the hot spot emission which indeed is
visible only  in the  phase ranges 0.25$\div$0.30  and 0.75$\div$0.85,
i.e.   when  it  is  seen  from the  ``outside''  and  the  ``inside''
respectively  (see Mason  et al.  2000). We also  note  that the
Balmer  emission lines  vary in  width and  intensity both  across the
orbit and  different orbital cycles (alternate spectra  in the trailed
spectrograms belong to two distinct orbital cycles). The same behavior
is not evident in the FeII lines  but we cannot say whether this is an
effect of the smaller S/N or  is rather and an intrinsic properties of
the system. 

On  the other side, the  HeII emission shows no  evidence for the
accretion disk contribution and produces  just a clear S-wave which is
signature of pure hot spot  emission.  This can be explained with the
fact that  the high  ionization emission lines  can form only  at high
temperatures, as in the hot spot region where the stream of in-falling
gas from the secondary star hits the outer edge of the accretion disk.

The  Balmer, FeII,  and  HeII lines  were  also used  to produce  back
projected  Doppler maps.   The maps  confirm our  previous conclusions
about the  emission lines  from the Balmer  series and  low ionization
elements. They mostly  form in the accretion disk  while the hot spot
fractional  contribution   is  small  and  washed  out   by  the  disk
emission\footnote{We may stress  that this was not at  all the case in
WZ~Sge where the hot spot emission  was contributing up to 50\% of the
line flux.}. The high ionization  emission line from HeII is confirmed
to form in the impact region within the accretion disk.

We plot  on top of  each Doppler map  the Roche lobe geometry  and the
stream trajectory  derived for H$\alpha$0242 (see  section 3.5). Also,
the circle drawn on top of the H and FeII Doppler maps fits by eye the
bulk of  the accretion  disk emission and  corresponds to 0.55  of the
white  dwarf Roche lobe  radius ($r_{L1}$).  In the  case of  the HeII
Doppler map we  draw the circle at 0.76 $r_{L1}$,  as derived from the
computation in Sec.~3.5.

   \begin{figure*}
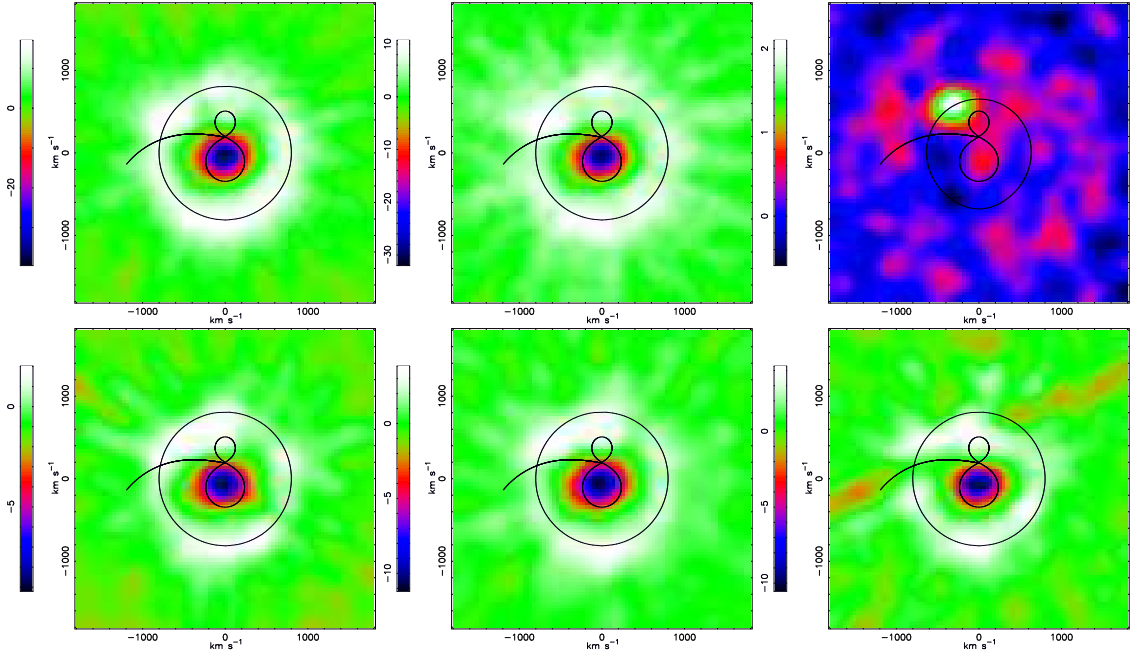

   \centering
   \rotatebox{-90}{\includegraphics[width=4.3cm,]{2349fig7.ps}}
   \rotatebox{-90}{\includegraphics[width=4.3cm,]{2349fig8.ps}}
   \rotatebox{-90}{\includegraphics[width=4.3cm,]{2349fig9.ps}}
   \rotatebox{-90}{\includegraphics[width=4.3cm,]{2349fig10.ps}}
   \rotatebox{-90}{\includegraphics[width=4.3cm,]{2349fig11.ps}}
   \rotatebox{-90}{\includegraphics[width=4.3cm,]{2349fig12.ps}}     
 \caption{Back  projected Doppler  maps: from  left to  right,  top to
 bottom:    H$\gamma$,   H$\beta$,   HeII    $\lambda$4686,   FeII(42)
 $\lambda\lambda$4923, 5018, and 5169. See text for details. }
         \label{f5}
   \end{figure*} 

\subsection{System parameters and geometry}

In the present  section we will make use  of the previous observations
and results to constraint the system geometry.

The continuum  light curve at  5500 \AA \  shows a 2 mag  deep eclipse
which is comparable to that  observed in Z~Cha and OY~Car (e.g. Ritter
and Kolb  2004). We  infer, to first  approximation, that  the orbital
inclination of  H$\alpha$0242 must be  similar to these  two eclipsing
systems, hence $i\sim$82$^o$.

Howell and Skidmore  (2002) present a $M_2$-$P$ relation  which can be
used to predict the mass of  the secondary star in both the hypothesis
of a  pre- and post-orbital  period minimum system. In  particular, we
found a secondary star of mass $M_2=0.17 M_\odot$ and radius $R_2=0.19
R_\odot$ in the  case of pre- orbital period  minimum system; while, a
secondary  star  of  mass  $M_2=0.03  M_\odot$  and  radius  $R_2=0.11
R_\odot$,  in case  of a  system which  has already  evolved  past the
orbital period minimum.

Knowing the orbital inclination and the mass of the secondary star, we
can solve  for $M_1$ (the white  dwarf mass), the  secondary star mass
function (i.e. equation 2.79 in Warner 1995). We derive a primary mass
of either 0.64 $M_\odot$ or $0.03 M_\odot$, depending on H$\alpha$0242
being a pre- or post-orbital period minimum system. The case $M_2=M_1=
0.03  M_\odot$  seems  unlikely   both  because  of  standard  stellar
evolution time scales (a star able to  form a low mass WD of just 0.03
$M_\odot$  has not  evolved off  the  main sequence,  yet, within  our
galaxy), and the standard  scenario for interacting binaries. Thus, we
discard it and conclude that H$\alpha$0242 has not reached the orbital
period minimum  yet and is likely to  have a white dwarf  of mass 0.64
$M_\odot$.

In the  case of $M_1=0.64 M_\odot$,  the mass ratio  will be $q=0.27$,
the binary separation is $\sim 4.8\times 10^{10}$ cm and the radius of
the  primary star Roche  lobe is  $2.4 \times  10^{10}$ cm.  Thus, the
accretion disk radius (see Sec.~3.3)  is $\sim 0.76$ times the size of
the Roche lobe, and 0.38 times the star separation $a$.

\subsection{The iron lines and the Balmer decrement}

As already  pointed out, FeII emission are  probably quite common
in CVs, though, they have never been studied carefully. In particular,
many  times they have  likely been  identified either  as HeI  (due to
limited spectral rage) or a combination  of FeII and HeI.  In the case
of H$\alpha$0242, we  see a multitude of FeII  emissions and just weak
HeI lines. Thus, it is reasonable  to conclude that (at least for this
system) the emission  lines at 4924\AA \ and  5018\AA \ consist mostly
of FeII transitions.

FeII emission in the optical were  observed since the 70s in Seyfert 1
galaxies (see  Osterbrock 1975 and Phillips 1977).  The two mechanisms
which  are believed  to produce  optical FeII  emissions in  Seyfert 1
galaxies are: {\it 1)} the resonance fluorescence (e.g. Wampler \& Oke
1967),  and {\it  2)} the  collisional excitation  (e.g.  Boksenber et
al. 1975).  In  the first case the UV photons emitted  by a hot source
($T_{eff}\leq25000$ K)  are absorbed by the iron  peak elements mostly
in the  wavelength range 2300$\div$2800\AA. These  UV absorption would
be followed  by downward  transitions in the  optical.  In  the second
case  the  UV  spectrum  should  be characterized  by  {\it  emission}
resonance lines in the UV region.
 
Now, the  optical spectrum of H$\alpha$0242 shows  emission lines from
FeII 42,  49, 37  and 38  similar to what  is seen  in many  Seyfert 1
galaxies.   We do  not  have UV  observation  of H$\alpha$0242,  still
several DN systems have been observed by HST, and for at least some of
them (OY Car, Horne et al.  1994; Z~Cha, WZ~Sge and V2051 Oph, Catalan
et al. 1998) the observations of an iron curtain (the UV absorption of
the iron peak  elements) has been reported.  It  is thus reasonable to
infer  that H$\alpha$0242 UV  spectrum is  characterized by  iron peak
absorptions,  and  that the  mechanism  responsible  for the  observed
optical FeII emission is indeed resonance fluorescence possibly from a
disk wind.  It would be  interesting, now, to observe/recover high S/N
spectra of DNe to search for  FeII emission in order to both derive the
fraction of DNe  which show optical FeII emission  lines and verify the
formation of these FeII  lines through the fluorescence mechanism.  We
are currently analyzing our database with such a purpose. Here, we can
comment that the  optical spectrum of WZ~Sge in  2002 was showing FeII
optical emissions (see  Fig.~4 of Howell et al.  2002), while the same
system was  not showing optical FeII  emission lines in  1996. This is
perfectly consistent  with the fact that WZ~Sge  underwent an outburst
in 2001  and the idea  that WD became  hotter as a consequence  of the
accretion,  exciting the  FeII in  an  increased disk  wind.  It  also
matches  the observation  by Catalan  et  al.  (1998)  who report  the
weakest  signature  of  iron   curtain  in  WZ~Sge  (before  the  2001
outburst). 

\begin{table}
\begin{center}
\scriptsize
\caption{The Balmer  decrement as derived from the  intensities of the
emission lines during times when they are not effected by the hot spot
emission.  The value  for the  WZ~Sge 1996  spectra is  from  Mason et
al. (2000).  The 2002 NTT WZ Sge  values (see Howell et  al. 2002) are
from a single spectrum covering  0.07 of the orbital period, while the
values for H$\alpha0242$ in 2002 (Howell  et al., 2002) span 0.2 of an
orbital period. The  intensity ratio of the blue and  red peak for the
2003 spectra  of H$\alpha$0242 (this  work) has been derived  from the
spectra within the phase ranges 0.02-0.11 and 0.43-0.56.}
\begin{tabular}{ccccc}
 object   &   date  &   H$\alpha$/H$\beta$   &  H$\gamma$/H$\beta$   &
H$\delta$/H$\beta$ \\
  & &  & & \\
  WZ~Sge & 1996  & 3.82  & - &  - \\
WZ~Sge &  08/2002 & 1.33  & 0.75 &  0.61 \\ 
H$\alpha$0242 &  08/2002 &
1.10 & 0.77 & 0.74 \\ 
H$\alpha$0242 & 11/2003 & - & 0.59 & - \\

\end{tabular}
\end{center}
\end{table}

In order to complete  our comparison between H$\alpha$0242 and WZ~Sge,
we  also compare  the Balmer  decrement  in the  two systems.   Direct
comparison is not simple/possible  because of the different wavelength
and  phase coverages\footnote{Some of  the data  are in  time resolved
mode, while others consist of  just a single spectrum.} of the spectra
in  our  hand. Still,  we  report  in  Table~3 the  Balmer  decrements
measured  in WZ~Sge  before the  outburst (spectra  from Mason  et al.
2000) and after the 2001 outburst  (spectrum from Howell et al.  2002)
with the Balmer decrement of H$\alpha$0242-28 measured in 2001 (Howell
et al. 2002)  and in this work.  From Table~3, it  is clear that: {\it
i)} the Balmer decrement in WZ~Sge before the 2001 outburst was larger
(and probably steeper)  than after its outburst; {\it  ii)} the Balmer
decrement  in H$\alpha$0242  is flatter  than in  WZ~Sge (at  least in
2002). We also note that in the 2003 spectra of H$\alpha$0242 there is
little or  no difference  in the Balmer  decrement resulting  from the
average spectrum  and/or the time-resolved spectra  (which exclude the
hot spot). This implies that  the opacity and probably the gas physics
within the disk and the hot  spot are similar.  The conclusion is that
the  line  forming  region  in  the accretion  disk  of  H$\alpha$0242
consists  of gas  which is  optically thicker,  thus,  probably denser
and/or warmer, than that in WZ~Sge.

\section{Summary and conclusions}

We have  presented time  series spectra of  the CV  H$\alpha$0242. The
object  was identified  as a  CV  candidate from  an H$\alpha$-R  band
survey (Davenhall et  al. 2001) and further suspected  as a candidate
TOAD by Howell et al. (2002).  We determined the system orbital period
of 107  min and showed evidence for  a very deep eclipse  in the light
curve (Sec.~3.1), both reported also  by Woudt et al. (2004). We infer
an orbital inclination of 82$^o$  from the observed eclipse depth.  We
measured  radial  velocities  of  the  Balmer  lines  and  derived  an
approximate   value   for   the   white   dwarf   Keplerian   velocity
(Sec.~3.3). We constrained the secondary  star on the basis of current
evolution  theory and  derived the  mass  of the  white dwarf  through
geometrical considerations (Sec.3.5).  We found $M_2=0.17 M_\odot$ and
$M_1=0.64 M_\odot$ for  the mass  of the secondary  star and the  white dwarf,
respectively.   We examined  the line  forming region  within the
accretion  disk (Sec.~3.4  and  3.6), observing  a  multitude of  FeII
emission  lines  and  a  flat  Balmer decrement.We  note  strong  FeII
emission  in  the  optical  spectrum  and postulate  that  many  other
cataclysmic variables show it as well, but previous observations often
lacked of either the proper wavelength coverage and/or sufficient S/N.
We  believe  that  the  FeII  lines  are  produced  by  the  resonance
fluorescence  mechanism and,  therefore, that  the white  dwarf  has a
relatively high  effective temperature.  From the Balmer  decrement we
also conclude that the gas both in the accretion disk and the hot spot
is  optically  thick.  The  hot  spot  was  found  to  not  contribute
significantly to both the Balmer and the FeII lines.  On the contrary,
we  show  evidence  for  pure  hot  spot emission  in  the  HeII  line
$\lambda$4684.  We interpret the HeII  emission as a signature of high
temperature in the impact region,  thus of high density gas within the
accretion disk.

The  results listed  above clearly  indicate that  H$\alpha$0242  is a
short orbital period system which has not yet evolved past the orbital
period   minimum.   The  accretion   disk   appearance  (see   trailed
spectrograms,  Doppler maps  and Table~3)  resemble that  of  a normal
SU~UMa star.

\begin{acknowledgements}
The authors wish to thank the ESO Director for the generous allocation
of time allowing these observations to be made.
      \end{acknowledgements}

\end{document}